\documentclass[aps,showpacs,nofootinbib]{revtex4}

\usepackage{graphicx}
\usepackage{graphics}
\usepackage{amssymb}
\usepackage{amsmath}
\usepackage{bm}

\newcommand{\half}{\textstyle{\frac{1}{2}}}
\newcommand{\bea}{\begin{eqnarray}}
\newcommand{\eea}{\end{eqnarray}}
\newcommand{\be}{\begin{equation}}
\newcommand{\ee}{\end{equation}}
\newcommand{\sig}{\sigma}

\begin{document}

\title{Extraction of proton form factors in the timelike region \\
from unpolarized $e^+e^- \to p\bar{p}$ events}

\author{Andrea Bianconi}
\email{andrea.bianconi@bs.infn.it}
\affiliation{
Dipartimento di Chimica e Fisica per i Materiali e per l'Ingegneria,  
via Valotti 9, 25100 Brescia, Italy, and\\
Istituto Nazionale di Fisica Nucleare, Sezione di Pavia, I-27100 Pavia, Italy}

\author{Barbara Pasquini}
\email{barbara.pasquini@pv.infn.it}
  
\author{Marco Radici}
\email{marco.radici@pv.infn.it}
\affiliation{
Dipartimento di Fisica Nucleare e Teorica, Universit\`{a} di Pavia, and\\
Istituto Nazionale di Fisica Nucleare, Sezione di Pavia, I-27100 Pavia, Italy}

\begin{abstract}
We have performed numerical simulations of the unpolarized $e^+ e^- \to p\bar{p}$ process 
in kinematic conditions under discussion for a possible upgrade of the existing DAFNE 
facility. By fitting the cross section angular distribution with a typical Born 
expression, we can extract information on the ratio $\vert G_E/G_M\vert$ of the proton 
electromagnetic form factors in the timelike region within a 5-10\% uncertainty. We have 
explored also non-Born contributions to the cross section by introducing a further 
component in the angular fit, which is related to two-photon exchange diagrams. 
We show that these corrections can be identified if larger than 5\% of the Born 
contribution, and if relative phases of the complex form factors do not produce severe 
cancellations.
\end{abstract}


\maketitle

%%%%%%%%%%%%%%%%%%%%%%%%%%%%%%%%%%%%%%%%%%%%%%%%%%%%%%%%%%%%

\section{Introduction}
\label{sec:intro}

The form factors of hadrons, as obtained in electromagnetic processes, provide
fundamental information on their internal structure, i.e. on the dynamics
of quarks and gluons in the nonperturbative confined regime. A lot of data for
nucleons have been accumulated in the spacelike region using elastic electron 
scattering (for a review, see Ref.~\cite{gao} and references therein). While the 
traditional Rosenbluth separation method suggests the well known scaling of the 
ratio $G_E/G_M$ of the electric to the magnetic Sachs form factor, new 
measurements on the electron-to-proton polarization transfer in 
$\vec{e}^{\,-} p \to e^- \vec{p}$ scattering reveal contradicting results, 
with a monotonically decreasing ratio for increasing momentum transfer 
$q^2 =-Q^2$~\cite{jlab}. This fact has stimulated a lot of theoretical work in 
order to test the reliability of the Born approximation underlying the Rosenbluth 
method (see Ref.~\cite{2gamma1,2gamma2,2gamma3} and references therein), and made 
it critical to deepen our knowledge of $G_E$ and $G_M$ also in the timelike 
region by mapping the $q^2$ dependence of their moduli and phases. 

Timelike form factors, as they can be explored in $e^+ e^- \to H \bar{H}$ or 
$p \bar{p} \to \ell^+ \ell^-$ processes, are complex because of the residual
interactions of the involved hadrons $H$ (protons $p$). Their absolute values can 
be extracted by combining the measurement of total cross sections and 
center-of-mass (c.m.) angular distributions of the final products. The phases are 
related to the polarization of the involved hadrons 
(see, e.g., Refs.~\cite{dubnick, brodsky1,egle1}). 

Experimental knowledge of the form factors in the timelike region is poor 
(for a review see, for example, Ref.~\cite{egle1}). There are no 
polarization measurements, hence the phases are unknown. 
The available unpolarized differential cross sections were integrated over a wide 
angular range, such that the relative weight of $\vert G_M\vert$ and 
$\vert G_E\vert$ is still unknown. In the data analysis, either 
the hypothesis $G_E=0$ or $|G_E|=|G_M|$ were used.  
While the former is arbitrary and certainly wrong near the physical threshold 
$q^2 = 4m^2$, with $m$ the nucleon mass, the latter is valid only at $q^2 = 4m^2$ 
but unjustified at larger $q^2$. As for the neutron, only one measurement is 
available by the FENICE collaboration~\cite{fenice} for $q^2 \leq 6$ 
GeV$^2$, which displays the same previous drawback. 

Nevertheless, these few data reveal very interesting properties.
The amplitudes in the timelike and spacelike regions are connected 
by dispersion relations~\cite{dr}. In particular, $|G_M|$ should
asymptotically become real and scale as in the spacelike region. However, a fit 
to the existing proton $|G_M|$ data for $q^2 \leq 20$ GeV$^2$ is compatible 
with a size twice as larger as the spacelike results~\cite{e685-2}. Moreover, 
the very recent data from the BaBar collaboration on $|G_E/G_M|$~\cite{babar} 
show that the ratio is surprisingly larger than 1, contradicting the 
spacelike results with the polarization transfer method~\cite{jlab} and the 
previous timelike data from LEAR~\cite{lear}. Also the few neutron data
for  $|G_M|$ are unexpectedly larger than the proton ones in the corresponding 
$q^2$ range~\cite{fenice}. Finally, all the available data show a steep rise of 
$|G_M|$ at $q^2 \sim 4m^2$, suggesting the possibility of interesting (resonant) 
structures in the unphysical region (for more details, see Ref.~\cite{baldini}). 

The possible upgrade of the existing DAFNE facility~\cite{LoI,roadmap} to enlarge the 
c.m. energy range from the $\phi$ mass to 2.5 GeV with a luminosity of at least 
$10^{32}$ cm$^{-2}$s$^{-1}$, would allow to explore with great precision 
the production of baryon-antibaryon pairs from the nucleon up to the $\Delta$. 

In the following, in Sec.~\ref{sec:formulae} we briefly review the formalism 
necessary to extract absolute values and phases of baryon timelike form factors 
from cross section data. 
In Sec.~\ref{sec:mc}, we outline the main features of our Monte Carlo 
simulation for the measurement of proton form factors. 
In Sec.~\ref{sec:reconstr}, we focus on unpolarized $p\bar{p}$ production, and in 
particular on the relevant problem of a precise extraction of the ratio 
$r_e \equiv \vert G_E/G_M \vert$ from the $\cos^2\theta$ term of the expected 
event distribution at any given $q^2$. The single-polarized case will be discussed 
elsewhere~\cite{next}. In Sec.~\ref{sec:twogamma}, we apply the 
same procedure to identify the presence of a contamination in the unpolarized 
event distribution by processes beyond the Born approximation, in particular
contributions from two-photon ($2\gamma$) exchange. Some concluding remarks are 
given in the final Sec.~\ref{sec:end}. 

%%%%%%%%%%%%%%%%%%%%%%%%%%%%%%%%%%%%%%%%%%%%%%%%%%%%%%%%%%%%%%%%%%%%%%%%%%%%%%%%

\section{General formalism} 
\label{sec:formulae}

The scattering amplitude for the reaction $e^+ e^- \to B \bar{B}$, 
where an electron and a positron with momenta $k_1$ and $k_2$, respectively, 
annihilate into a spin-$\half$ baryon and an antibaryon with momenta $p_1$ and 
$p_2$, respectively, is related by crossing to the corresponding scattering 
amplitude for elastic $e^- B$ scattering. In the timelike case, and in presence 
of a possible $2\gamma$ exchange term, quite a few independent real scalar 
functions are required to describe the process. There are several equivalent 
representations of them; here, we use the one involving the axial 
current following the scheme of Ref.~\cite{egle2}. The scattering amplitude
can be fully parametrized in terms of three complex form factors: 
$G_E(q^2,t), \, G_M(q^2,t),$ and $G_A(q^2,t)$, which are functions of 
$q^2=(k_1+k_2)^2$ and $t=(k_2-p_1)^2$. They refer to an identified 
baryon-antibaryon pair in the final state, rather than to an identified isospin 
state. In the following, we will use $\cos\theta$ instead of the variable $t$, 
with $\theta$ the angle between the momenta of the positron and the recoil proton 
in the c.m. frame. 

In the Born approximation, $G_E$ and $G_M$ reduce to the usual Sachs form factors 
and they do not depend on $\cos\theta$, while $G_A=0$. For the construction of 
our Monte Carlo event generator, we rely on the general and extensive 
relations of Ref.~\cite{egle2} up to the single polarization case. 
These relations assume small non-Born terms and include them up to order 
$\alpha^3$ (with $\alpha$ the fine structure constant), i.e. they consider 
$2\gamma$ exchange only via their interference with the Born amplitude. 
These corrections introduce explicitly six new functions, that all can depend on
$\cos\theta$: the real and imaginary parts of $\Delta G_E$ and $\Delta G_M$, i.e. 
the $2\gamma$ corrections to the Born magnetic and electric form factors, and the 
real and imaginary parts of the axial form factor $G_A$. 
The role of these corrections in the unpolarized  and single polarized
cross sections is very simple: it can be deduced from the Born term
by adding the  contribution of   $G_A$ and
substituting the Born form factors with the 
"$2\gamma-$improved" form factors, i.e.   $G_{E,M}(q^2)$ $\rightarrow$ 
$G_{E,M}(q^2)+\Delta G_{E,M}(q^2,cos\theta).$ In Ref.~\cite{egle2}, this fact 
is somehow hidden by neglecting terms of order $\alpha^4$. However,
for sake of simplicity we keep $2\gamma$ effects 
via the axial form factor only. 
Within this scheme, when summing events with positive and negative polarization, 
the unpolarized cross section can be written as
\bea
\frac{d\sig^o}{d\cos\theta} &= 
& a(q^2) \, [ 1+ R(q^2) \, \cos^2 \theta ] - b(q^2) \, \mathrm{Re}
[G_M(q^2)\,{G_A}^\ast(q^2,\cos \theta)]\, \cos \theta \; , \label{eq:unpolxsect} 
\\
a(q^2) &= &\frac{\alpha^2 \pi}{2q^2}\,\frac{1}{\tau}\,\sqrt{1-\frac{1}{\tau}}\,
\left( \tau |G_M|^2 + |G_E|^2\right) \; , \quad 
b(q^2) = \frac{2\pi \alpha^2}{q^2}\,\frac{\tau -1}{\tau} \; ,\label{eq:ab} \\
R(q^2) &= &\frac{\tau |G_M(q^2)|^2-|G_E(q^2)|^2}{\tau |G_M(q^2)|^2+|G_E(q^2)|^2} 
\; , \quad \tau = \frac{q^2}{4m^2} \; . \label{eq:rtau}
\eea
Measurements of the unpolarized cross section at fixed $q^2$ for different values 
of $\theta$ allow us to fit the different $\cos^n\theta$ terms, from which we can 
extract $\vert G_E/G_M\vert$ and some information on $G_A(q^2,cos\theta)$.

For spin-$\half$ baryons with polarization $\mathbf{S}_B$, the cross section is
linear in the spin variables, i.e. $d\sig = d\sig^o \, (1+ \mathcal{P} \, 
\mathcal{A} )$, with $d\sig^o$ from Eq.~(\ref{eq:unpolxsect}) and
$\mathcal{A}$ the analyzing power. In the c.m. frame, three 
polarization states are observable~\cite{dubnick,brodsky1}: the longitudinal 
$\mathcal{P}_z$, the sideways $\mathcal{P}_x$, and the normal $\mathcal{P}_y$. 
The first two ones lie in the scattering plane, while the normal points in the 
$\mathbf{p}_1 \times \mathbf{k}_2$ direction, the $x, y, z,$ forming a 
right-handed coordinate system with the longitudinal $z$ direction along
the momentum of the outgoing baryon. The $\mathcal{P}_y$ is particularly interesting,  
since it is the only observable that does not require a polarization in the 
initial state~\cite{dubnick,brodsky1}. With the above approximations, it can be
deduced by the spin asymmetry between events with positive and negative
normal polarizations:
\be
\mathcal{P}_y = \frac{1}{\mathcal{A}_y}\, 
\frac{d\sig^\uparrow - d\sig^\downarrow}{d\sig^\uparrow + d\sig^\downarrow} 
= \frac{b(q^2)}{2\sqrt{\tau -1}\, d\sig^o}\, \sin \theta \, 
\Big\{ \cos \theta \, \mathrm{Im}\left[ G_M(q^2) \, G_E^\ast(q^2) \right]
 - \sqrt{\frac{\tau -1}{\tau}}\, \mathrm{Im}\left[ G_E(q^2) \, 
{G_A}^\ast(q^2,\cos \theta) 
\right] \Big\} \; .
\label{eq:py}
\ee
This spin asymmetry can be nonvanishing even without polarized lepton beams,
because it is produced by the mechanism $\mathbf{p}_1 \times \mathbf{k}_2
\cdot \mathbf{S}_B$, which is forbidden in the Born approximation
for the spacelike elastic scattering~\cite{brodsky1}. However, the measurement of 
$\mathcal{P}_y$ alone does not completely determine the phase difference of the 
complex form factors. By defining with $\delta_E$ and $\delta_M$ the phases of the 
electric and magnetic form factors, respectively, the Born contribution is 
proportional to $\sin (\delta_M - \delta_E)$, leaving the ambiguity between 
$(\delta_M - \delta_E)$ and $\pi - (\delta_M - \delta_E)$. Only
the further measurement of $\mathcal{P}_x$ can solve the problem, because 
$\mathcal{P}_x \propto \mbox{\rm Re}(G_M^{} \,G_E^\ast) \propto \cos (\delta_M -
\delta_E)$~\cite{brodsky1}. But at the price of requiring a polarized electron 
beam.

%%%%%%%%%%%%%%%%%%%%%%%%%%%%%%%%%%%%%%%%%%%%%%%%%%%%%%%%%%%%%%%%%%%%%%%%%%%%%%%

\section{General features of the numerical simulation}
\label{sec:mc}

We consider the $e^+ e^- \to p\bar{p}$ process. Events are generated in the 
variables $q^2, \, \theta ,\, \phi$ (the azimuthal angle of the proton momentum
with respect to the scattering plane), and $S_y$ (the proton polarization normal
to the scattering plane). Then, the distribution is summed upon the spin and
integrated upon $\phi$. 

The exchanged timelike $q^2$ is fixed by the beam energy. Only the scattering 
angle $\theta$ is randomly distributed. For a given $q^2$, the Born term is 
responsible for the $[1+R(q^2)\,\cos^2 \theta ]$ behaviour of the event 
distribution. Since at this level and with the discussed approximations no other 
dependence in $\theta$ is present, observed systematic deviations from the 
$[1 + R(q^2)\, \cos^2\theta ]$ behaviour will be interpreted as a clear signature 
of non-Born terms. 

An overall sample of $300\,000$ events has been considered with 
$3.8 \leq q^2 \leq 6.2$ GeV$^2$ and $\vert \cos\theta \vert < 0.9$. Since the 
integrated cross section for $e^+ e^- \to p \bar{p}$ in the considered region 
is approximately 1 nb, at the foreseen luminosity of $10^{32}$ cm$^{-2}$s$^{-1}$ 
this sample can be collected in one month with efficiency 1. The upper $q^2$ 
cutoff is consistent with the upper limit of the c.m. energy in the presently
discussed upgrade of DAFNE~\cite{LoI,roadmap}. The lower limit includes the $p\bar{p}$ 
threshold, but in our simulation we avoid this region which is characterized by 
peculiar mechanisms like, e.g., the Coulomb focussing and possible subthreshold 
resonances. 

Results depend much on the binning of the events. In fact, while a larger number 
of bins in the same $(q^2,\theta)$ region implies a more precise determination 
of $q^2$ and $\theta$, at the same time it 
leaves each bin with much less events and, consequently, with larger error bars. 
We find the best compromise with 6 equally spaced $q^2$ bins with width 
$\Delta q^2 = 0.4$ GeV$^2$, and 7 equally spaced $\cos\theta$ bins with width 
$\Delta \cos\theta \approx 0.257$ for the solid angle 
$\vert \cos\theta\vert < 0.9$. Incidentally, we notice that with 
$\Delta q^2 = 0.4$ GeV$^2$ the deviation inside each bin between the statistical 
average $q^2$, calculated in the simulation, and the central value 
$(q^2_{max}-q^2_{min})/2$ is about 3\%. 

Since it is well known that $\vert G_M(q^2)\vert \sim 1/q^4$~\cite{brodsky}, from
Eqs.~(\ref{eq:unpolxsect}-\ref{eq:rtau}) we expect the cross section to
approximately fall like $1/q^{10}$. This implies that bins at higher $q^2$ are
scarcely populated. For example, out of the $300\,000$ selected events only 700 
are accumulated in the $5.8\leq q^2 \leq 6.2$ GeV$^2$ range for each $\cos\theta$ 
bin. In these conditions, a gaussian fluctuation of $1\sigma$ differs by 5\% from 
the average value; therefore, deviations of at least 10\% from the Born cross 
section are needed to be recognized as systematic effects due to non-Born terms. 
The situation is evidently more favourable at lower $q^2$: in the 
$3.8\leq q^2 \leq 4.2$ GeV$^2$ range around $23\,000$ events are accumulated in
each $\cos\theta$ bin. Of course, it could be possible to asymmetrically split the
beam time in order to make the statistics of each $q^2$ bin more homogeneous. More
importantly, since in a $e^+ e^-$ collider experiment $q^2$ is fixed by the
colliding particle energy, it could be more convenient to concentrate on narrow
subranges like, e.g., $4\leq q^2 \leq 4.1$ GeV$^2$ and $4.4 \leq q^2 \leq 4.5$
GeV$^2$. The reconstruction efficiency does not change much when the same number 
of events is concentrated in two bins centered around the same $q^2$ value but 
with different widths $\Delta q^2$, at least for $\Delta q^2 \ll 1$ GeV$^2$. We
define as "default conditions" the ones where the whole possible range in $q^2$ is
divided in equally spaced bins and each one is sampled with the same beam time; our
results are presented in this framework. A discussion about other possible choices
is beyond the scope of this paper.

Events can be generated only by inserting specific parametrizations of the proton
form factors in the cross section. For $G_E$ and $G_M$, several models can be
considered in the timelike region~\cite{brodsky1,egle1}, mostly derived from 
extrapolations from the spacelike region. 
%Models based on dispersion
%relations~\cite{dr} cannot be conveniently used in our Monte Carlo simulation,
%where the cross section needs to be calculated more than $10^6$ times to extract
%$300\,000$ events. 
We selected the parametrizations of Refs.~\cite{iachello,lomon}, because they have 
been recently updated in Ref.~\cite{egle1} by simultaneously fitting both the 
spacelike and timelike available data. 
Moreover, both cases release separate parametrizations for the
real and imaginary parts of $G_E$ and $G_M$. We indicate the former as the IJLW
parametrization (from the initials of their authors), and the latter as the Lomon
parametrization. 

Since the explored $q^2$ range is not large, it is reasonable to make also much
simpler, but equally effective, choices. Accordingly, we use also a parametrization
where all form factors are real and proportional to the same dipole term 
$1 / (1 + q^2/q_o^2)^2$, with $q_o = 0.71$ GeV, as in the spacelike region. This 
choice looks
reasonable also for $G_A$, of which nothing can be said except that hopefully its
modulus could have the same dipole trend in the considered $q^2$ range. All form
factors being proportional to the same function, the actual parameters are the
ratios $r_e = \vert G_E/G_M \vert$ and $r_a = \vert G_A/G_M\vert$, and the relative
phases $\beta_e$ and $\beta_a$ of $G_E$ and $G_A$, respectively, with respect to
$G_M$; they determine the sign of the corresponding ratios. We will refer to the
parametrization Dip0 when $r_e = r_a = 0$; to Dip1 when $r_e=1$ and $r_a=0$; to
Dip$2\gamma$ when $r_e=1$ and $r_a=0.2$. Since the Born term in the cross section
depends only on $\vert G_E \vert^2$ and $\vert G_M \vert^2$, we can safely take
$\beta_e = 0$. We refer the reader to Sec.~\ref{sec:twogamma} for a discussion
about $r_a$ and $\beta_a$.

%%%%%%%%%%%%%%%%% FIG. 1
\begin{figure}[h]
\centering
\includegraphics[width=8cm]{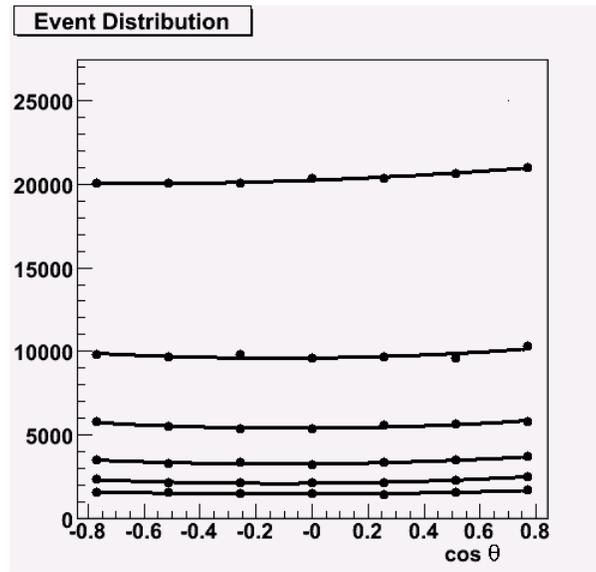}  
\caption{Angular distribution of $300\,000$ events for the $e^+ e^- \to p\bar{p}$
process at $3.8\leq q^2 \leq 6.2$ GeV$^2$ and $\vert \cos\theta \vert < 0.9$, and
with the Dip1 parametrization of proton form factors (see text). The points
correspond to 7 equally spaced $\cos\theta$ bins for each of 6 equally spaced
$q^2$ bins. Solid curves are the results of a 3 parameters angular fit with
$B\propto r_a =\vert G_A/G_M\vert = 0$ (see text). The
highest curve corresponds to the lowest bin $3.8 \leq q^2 \leq 4.2$ GeV$^2$; next
lower curve to the adjacent bin $4.2 \leq q^2 \leq 4.6$ GeV$^2$, and so on.}
\label{fit11000}
\end{figure}

%%%%%%%%%%%%%%%%%%%%%%%%%%%%%%%%%%%%%%%%%%%%%%%%%%%%%%%%%%%%%%%%%%%%%%

\section{Reconstruction of $\vert G_E/G_M \vert$}
\label{sec:reconstr}

As already stressed in the Introduction, one of the puzzling features of the
available data is that the BaBar collaboration reports a value for 
$\vert G_E/G_M\vert$ bigger than 1 in the timelike region~\cite{babar}, which
contradicts the spacelike results~\cite{jlab} and the previous timelike data from
LEAR~\cite{lear}. Therefore, it is important to explore the possibility of new and
more precise measurements of $\vert G_E/G_M \vert$ in the unpolarized 
$e^+ e^- \to p\bar{p}$ process at the upgraded DAFNE~\cite{LoI,roadmap}.

Our analysis consists of three steps; for sake of simplicity, we describe it in
the following for the case of the parametrization Dip1:
\begin{itemize}
\item[1)] We generate $300\,000$ events for $3.8 \leq q^2 \leq 6.2$ GeV$^2$ and 
$\vert \cos\theta\vert < 0.9$ (i.e., for 
$25^{\mathrm{o}} \leq \theta \leq 155^{\mathrm{o}}$). The sorted events are 
divided into 6 equally spaced $q^2$ bins of width $\Delta q^2 = 0.4$ GeV$^2$. In 
each $q^2$ bin, the events are divided into 7 equally spaced $\cos\theta$ bins 
with width $\Delta \cos\theta \approx 0.257$. In Fig.~\ref{fit11000}, the points
show the 6 distributions in $q^2$, each one consisting of 7 points describing the
distribution in $\cos\theta$. From the discussion in the previous section, the
highest $\cos\theta$ distribution corresponds to the lowest bin 
$3.8 \leq q^2 \leq 4.2$ GeV$^2$; the next lower distribution to the adjacent bin
$4.2 \leq q^2 \leq 4.6$ GeV$^2$, and so on.

%%%%%%%%%%%%%%%%%%% FIG. 2
\begin{figure}[h]
\centering
\includegraphics[width=8cm]{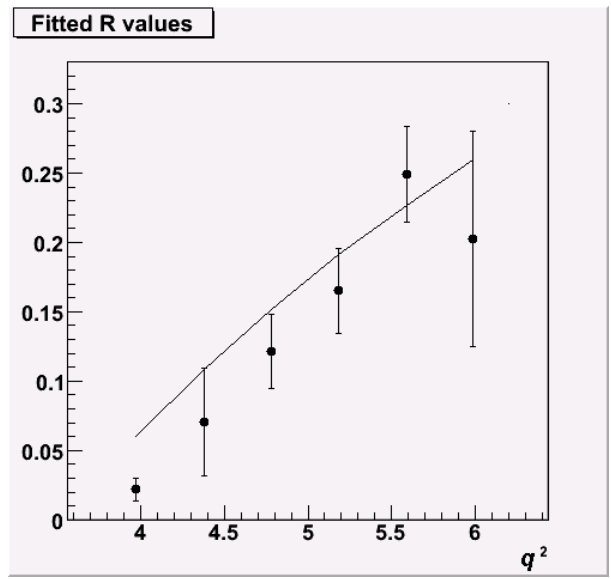}  
\caption{Fitted $R$ values from the event distribution of 
Fig.~\protect{\ref{fit11000}}, according to Eq.~(\protect{\ref{eq:fitting}}) (see
text). Solid line represents the expected values for the Dip1 parametrization
according to Eq.~(\protect{\ref{eq:re_rel}}) (see text).}
\label{R11000}
\end{figure}
%%%%%%%%%%%%%%%%%

\item[2)] For each $q^2$ bin, the $\cos\theta$ dependence of the event 
distribution is fitted by a function of the form 
\begin{equation}
N(\cos\theta) =\ A\Big(1\ -\ B\, \cos\theta\ +\ R \, \cos^2\theta\Big)\;, 
\label{eq:fitting}
\end{equation}
with three fitting parameters $A, B$, and $R$. The results of each fit are
represented by the solid curves in Fig.~\ref{fit11000}. The parameter $A$ is not 
crucial for the following discussion and it will not be considered. The parameter 
$B$ contains the effects of $2\gamma$ exchange which will be discussed in 
Sec.~\ref{sec:twogamma}. Here, we just notice that, with the input 
$r_a = \vert G_A/G_M\vert = 0$ and discarding $2\gamma$ effects in $G_E$ and $G_M$, 
the simulation produces almost perfectly symmetric
curves in $\cos\theta$, indicating that within the statistical uncertainty 
$B\sim 0$, as expected. The parameter $R$ must obviously be identified with the
function $R(q^2)$ of Eq.~(\ref{eq:rtau}), and it is related to 
$r_e =\vert G_E/G_M \vert$ by  
\begin{equation}
R\ =\ \frac{\tau-r_e}{\tau+r_e} \; .  
\label{eq:re_rel} 
\end{equation}
The fitted $R$ values are reported 
in Fig.~\ref{R11000} against the corresponding average $q^2$ values of each bin. 
The error bars are the output of the fitting procedure, i.e. they represent the
uncertainty in determining $R$ when using the form~(\ref{eq:fitting}) to fit the
event distribution in Fig.~\ref{fit11000}. The solid curve in 
Fig.~\ref{R11000} represents the expected value of $R$ when the Dip1 
parametrization for the form factors is inserted in Eq.~(\ref{eq:re_rel}) through
$r_e$.

%%%%%%%%%%%%%%%%% FIG. 3
\begin{figure}[h]
\centering
\includegraphics[width=8cm]{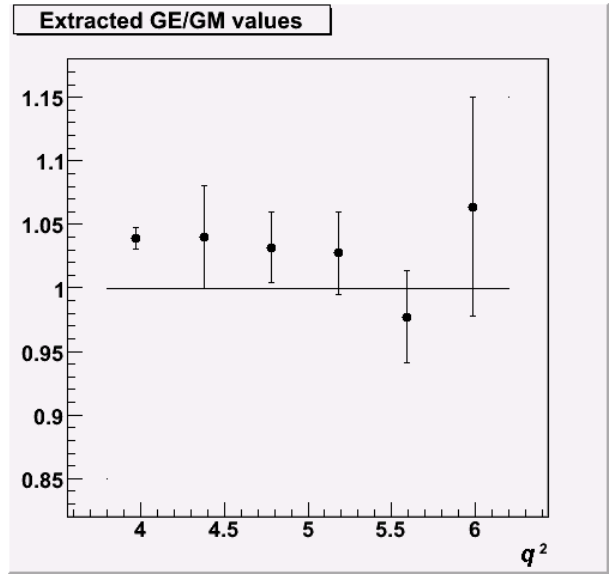}  
\caption{
Extracted $r_e = \vert G_E/G_M\vert$ values from the event distribution of 
Fig.~\protect{\ref{fit11000}}. Solid line gives the expected value according to
the Dip1 parametrization (see text).} 
\label{EM11000}
\end{figure}
%%%%%%%%%%%%%%%%

\item[3)] If we define $\Delta R$ as half of the error bar associated with a given
$R$ value in Fig.~\ref{R11000}, we can also define the pair 
$(R_{min} = R-\Delta R,\, R_{max} = R+\Delta R)$. By inverting the
relation~(\ref{eq:re_rel}), it is possible also to deduce the corresponding pair
$(r_{e_{min}}, \, r_{e_{max}})$, which can be transformed into $(r_e, \Delta r_e)$
with $\Delta r_e = (r_{e_{max}} - r_{e_{min}})/2$. These extracted values of 
$r_e = \vert G_E/G_M\vert$ and the associated error bars $\Delta r_e$ are
displayed for each corresponding $q^2$ bin in  Fig.~\ref{EM11000}. The solid line 
represents the expected $r_e$ value (in this case, $r_e = 1$ because we are using
the Dip1 parametrization). The comparison between expected and reconstructed 
$r_e$ values displays the potential reliability of the measurement of $r_e$ with 
a sample of $300\,000$ events and with the chosen binning in $q^2$. 
\end{itemize}

Our reconstruction procedure is based on the structure of the cross
section~(\ref{eq:unpolxsect}), which naturally suggests to fit the 
directly observable parameter $R$ and then to link it to 
$r_e$ 
according to Eq.~(\ref{eq:re_rel}). It must be remarked that the
usual formula for the error propagation cannot be used to derive $\Delta r_e$ from
$\Delta R$, because for some $R$ values the relative error is very large (see
Fig.~\ref{R11000}). We stress that the error bars are related to the fitting
procedure, i.e. they give the integral deviation between the test
function~(\ref{eq:fitting}) and the data points. They are not directly related to
the statistical errors on the average population of each bin. In the measurement 
of a really random variable, in the limit of infinite number of events, the 
statistical error on the population of a single bin tends to vanish because of 
the central limit theorem. In the same limit, the error from the fitting 
procedure does not tend to vanish unless the number of parameters is equal to the 
number of bins (in this case, 7 $\cos\theta$ bins for each independent fitting 
procedure). Anyway, the statistical fluctuations can be estimated from the 
distance between expected and reconstructed $r_e$ values. 

In order to extract $r_e$ from $R$ we need to assign a common $\tau$ value, i.e. 
$q^2$, to all the events falling into a specific $q^2$ bin. Integrating over all 
the generated events, we have calculated the average $q^2$ for each $q^2$ bin, 
and used it for the extraction of $r_e$. We recall that this average $q^2$ 
differs by about 3\% from the central value $(q^2_{max}-q^2_{min})/2$ deduced from
the bin width $\Delta q^2 = q^2_{max} - q^2_{min}$.

It is important to notice also that, despite the large number of available 
events, the region with the smallest $q^2$ is delicate. In fact, $R$ is small 
but has not a small derivative in $q^2$ near threshold. This reflects in a 
striking contrast between the very small fitting error bar, on one side, and 
a worse comparison between reconstructed and expected $r_e$ than at larger $q^2$, 
on the other side. 

%%%%%%%%%%%%%%%%% FIG. 4
\begin{figure}[h]
\centering
\includegraphics[width=8cm]{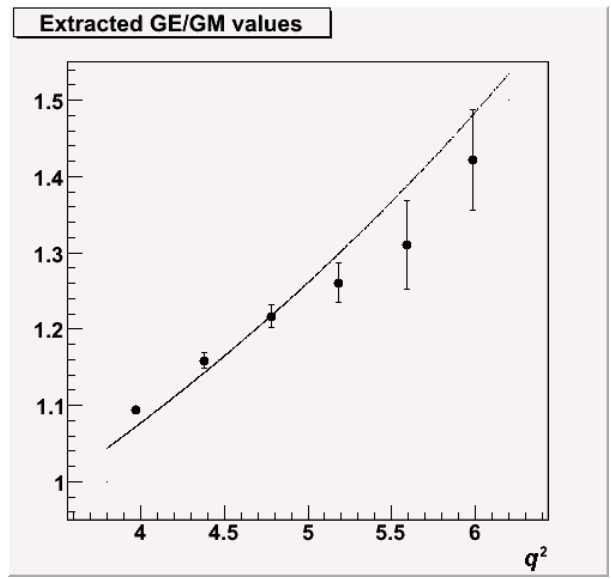}   
\caption{Extracted $r_e = \vert G_E/G_M\vert$ values from the event distribution 
in the same conditions as in Fig.~\protect{\ref{fit11000}} but for the IJLW
parametrization of Ref.~\protect{\cite{iachello}}, as updated in
Ref.~\protect{\cite{egle1}} (see text). Solid line is the 
expected result.}
\label{EMiachello}
\end{figure}
%%%%%%%%%%%%%%%%%%%%%%%%%%%%%

We have repeated the reconstruction procedure also for the Dip0 parametrization,
i.e. for $r_e = \vert G_E/G_M\vert = 0$, which has been used in some experiments
to extract the timelike $\vert G_M\vert$ data from unpolarized cross section
measurements. The comparison between expected and reconstructed $r_e$ is more
difficult, because not only the fitting error is obviously larger, but also the
statistical one. However, since the considered $q^2$ are rather close to the 
physical threshold, $\vert G_E/G_M \vert$ may not be very different from its
threshold value of unity. According to the results with the Dip1 parametrization, 
this should simplify the task of a precise extraction of $r_e$ in a real 
experiment.

%%%%%%%%%%%%%%%%% FIG. 5
\begin{figure}[h]
\centering
\includegraphics[width=8cm]{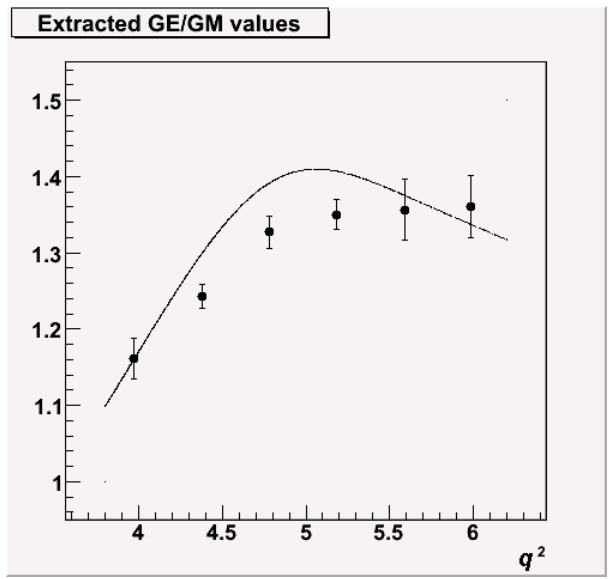}  
\caption{Extracted $r_e = \vert G_E/G_M\vert$ values from the event distribution
in the same conditions as in Fig.~\protect{\ref{fit11000}} but for the Lomon
parametrization of Ref.~\protect{\cite{lomon}}, as updated in
Ref.~\protect{\cite{egle1}} (see text). Solid line is the expected 
result.}
\label{EMlomon}
\end{figure}
%%%%%%%%%%%%%%%%%

To verify this conjecture, we have repeated the procedure starting from the 
realistic parametrizations IJLW (Fig.~\ref{EMiachello}) and Lomon 
(Fig.~\ref{EMlomon}). Again, the solid curves show the expected values of $r_e$ 
as calculated for each different choice. 

Three peculiar features distinguish these two examples from the previous 
ones: $G_E$ and $G_M$ are complex objects; $r_e$ is not a constant; it is 
close to 1 but bigger than 1, in agreement with the recent findings of the BaBar
experiment~\cite{babar}. In both cases, the quality of the reconstruction of 
$r_e$ is very good, confirming that in most of the $q^2$ bins the extraction of 
$\vert G_E/G_M\vert$ is possible with an accuracy within 10\% for the considered 
sample and binning. This is mainly due to the simple dependence of the 
$\cos^2\theta$ term in the angular distribution upon only one parameter, i.e. 
$r_e = \vert G_E/G_M\vert$; to the neglect of more complicated angular terms, like
$\cos^4 \theta$; and to the vicinity of $q^2$ to the physical threshold
$q^2=4m^2$, which implies $\vert G_E/G_M\vert \sim 1$.

%%%%%%%%%%%%%%%%%%%%%%%%%%%%%%%%%%%%%%%%%%%%%%%%%%%%%%%%%%%%%%%%%%%%%%%5

\section{The two-photon contribution}
\label{sec:twogamma}

As already anticipated in Sec.~\ref{sec:formulae}, we account for non-Born
contributions to the unpolarized cross section only through the axial form factor
$G_A$. According to the Dip$2\gamma$ parametrization discussed in
Sec.~\ref{sec:reconstr}, $G_A$ is modeled as
\begin{equation}
G_A(q^2,\cos\theta)\ =\ r_a \, e^{i\beta_a\pi} \, G_M(q^2) \; ,
\label{eq:par_ga}
\end{equation}
where we neglect any explicit dependence on $\cos\theta$. We take $G_M$ real and
with the dipole form. Indeed, in the kinematical region under analysis the ratio 
$G_E/G_M$ is complex, but most likely dominated by its real part. A nonvanishing 
phase is due to the reinteraction between the two hadrons in the final state. 
Very close to the threshold (i.e., in a relative 
$s$ wave) the phenomenon presents peculiar features, but from the onset of 
the $p$ wave contribution onwards, an approximately semiclassical black-disk 
regime takes over~\cite{pbar1}. In these conditions, the main effect of 
the $p\bar{p}$ interaction is a flux damping. 
The transition from the near-threshold to the black-disk regime is 
signalled by the change of sign of the $\rho$ parameter, i.e. the 
ratio between the real and imaginary parts of the forward 
$p \bar{p}$ scattering amplitude. Such a transition takes place below the $q^2$ 
range considered in our analysis. In the black-disk regime, it is natural to 
think that the flux absorption leads to the appearance of an imaginary part for 
$G_E$ and $G_M$. However, the absorption cross section is much below the 
unitarity limit, suggesting that the phases of $G_E$ and $G_M$ may be small. 
For $G_A$ it is not possible to conclude that it is "quasi-real". 
$\vert G_A\vert$ is presumably small, and the cuts on the two-photon lines 
involving 
all the possible on-shell intermediate states in the amplitude, may produce a 
relevant imaginary part. 

The $G_A$ form factor enters the unpolarized cross section 
with a contribution proportional to $\cos\theta\,\mathrm{Re}(G_M^{} G_A^*)$. 
Therefore, we can have
\be
\mathrm{Re} [G_M^{}\, G_A^*] \approx G_M\, \mathrm{Re} [G_A^*] 
\approx r_a \, \cos \beta_a \pi \, G_M^2 \equiv rr_a \, G_M^2 \; .
\label{eq:rra}
\ee
In our simulation with the Dip$2\gamma$ parametrization, we consider 
$G_M,\, G_E$, and $G_A$, all real functions of $q^2$, with 
$r_e = \vert G_E/G_M\vert = 1$ and $rr_a \approx r_a = \vert G_A/G_M\vert = 0.2$, 
i.e. with $\beta_a = 0$. This is the most optimistic situation we may guess, 
taking into account that: (i) no realistic models are available for $G_A$; (ii) 
in the spacelike case, the assumption that $r_a$ is at most 0.2 does not 
contradict the data; (iii) in the timelike case, $G_A$ should have a nonvanishing
phase.  

%%%%%%%%%%%%%%%%% FIG. 6
\begin{figure}[h]
\centering
\includegraphics[width=8cm]{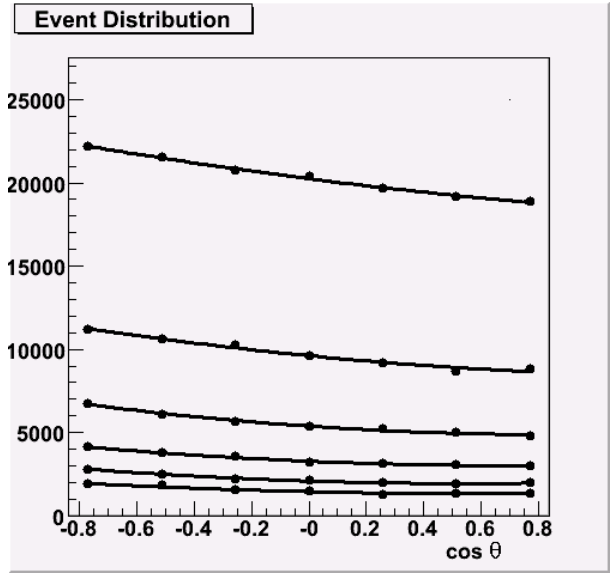} 
\caption{Angular distributions for $300\,000$ events for the $e^+ e^- \to
p\bar{p}$ process in the same conditions as in Fig.~\protect{\ref{fit11000}}, but
for $r_a=\vert G_A/G_M\vert = 0.2$. }
\label{fit11200}
\end{figure}
%%%%%%%%%%%%%

In Fig.~\ref{fit11200}, the generated events are sorted in $q^2$ and $\cos\theta$
bins as in Fig.~\ref{fit11000}. We can see now that $2\gamma$ effects show up in a
marked left-right asymmetry of points in $\cos\theta$. The solid curves are 
obtained from the fitting formula~(\ref{eq:fitting}), where now the parameter $B$ 
is given by
\begin{equation}
B(q^2)\ =\ r_a \, f(q^2,r_e)\; , 
\label{eq:bparam}
\end{equation}
with
\begin{equation}
f(q^2,r_e)\ = 4 \frac{\sqrt{\tau(\tau-1)}}{\tau + r_e^2} \; .
\label{eq:bparam2} 
\end{equation}

%%%%%%%%%%%%%%%%% FIG. 7
\begin{figure}[h]
\centering
\includegraphics[width=8cm]{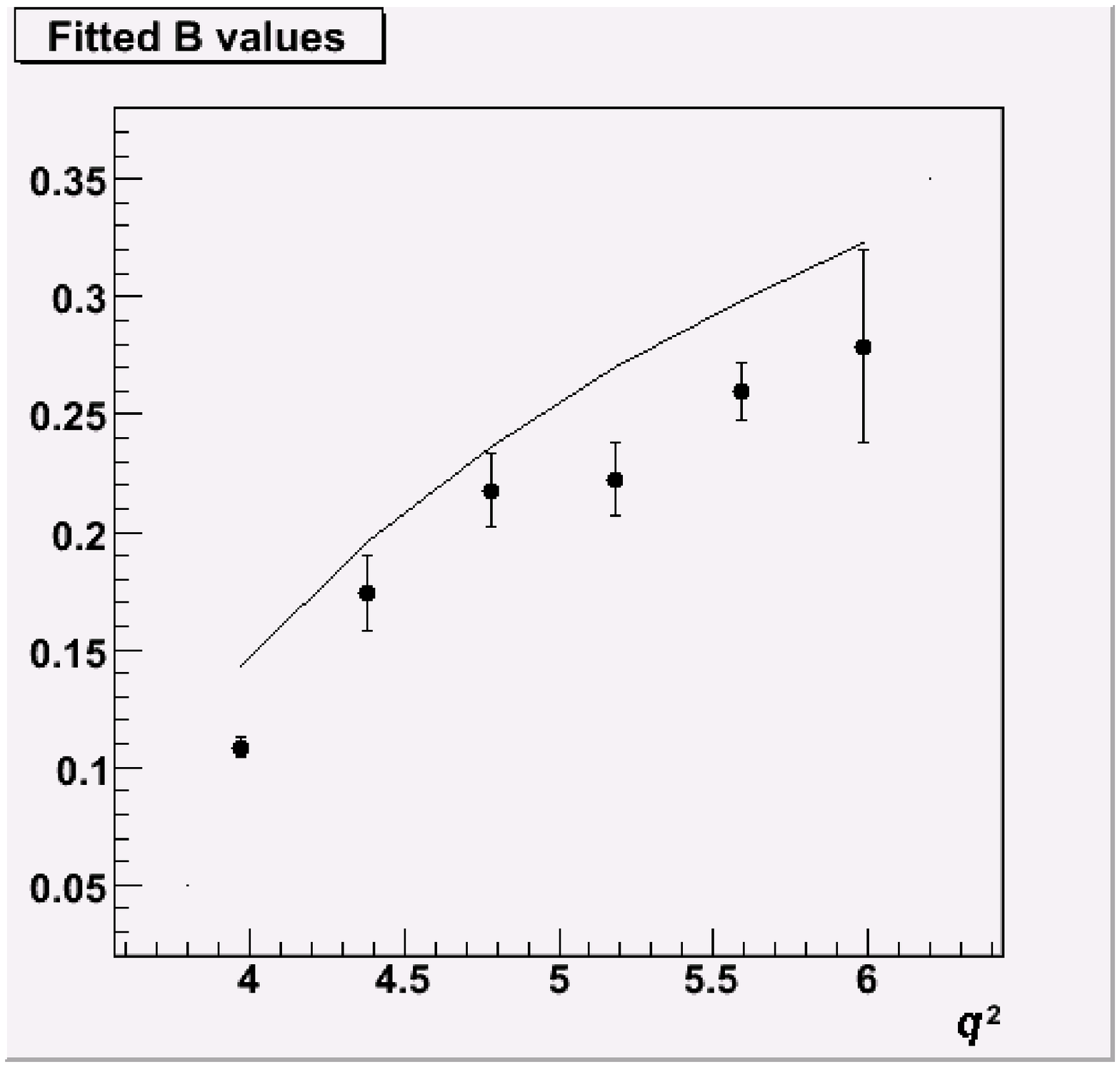}  
\caption{
Fitted $B$ values from the event distributions of Fig.~\protect{\ref{fit11200}}
according to Eq.~(\protect{\ref{eq:fitting}}) (see text). Solid line represents
the expected value for the Dip$2\gamma$ parametrization according to
Eq.~(\protect{\ref{eq:bparam}}) (see text).}
\label{B11200}
\end{figure}
%%%%%%%%%%%%%%

In Fig.~\ref{B11200}, the fitted values of $B$ and the relative errors are shown. 
As usual, the solid line shows the expected result, by inserting $r_e=1$ and
$r_a=0.2$ in Eqs.~(\ref{eq:bparam}-\ref{eq:bparam2}). 

In order to estimate  $r_a$ from $B$, we assume small errors for both $B$ and 
$r_e$, with the caveat that such an assumption is not valid near the lowest $q^2$ 
threshold. Using the formula for the systematic error propagation, we get
\begin{equation}
\Delta r_a\ =\ \frac{r_a}{B} \, \left[
\Delta B + \ \frac{2 r_e}{\tau + r_e^2}\, B \, \Delta r_e
 \right] \; .  
\label{eq:err_prop}
\end{equation}

%%%%%%%%%%%%%%%%% FIG. 8
\begin{figure}[h]
\centering
\includegraphics[width=8cm]{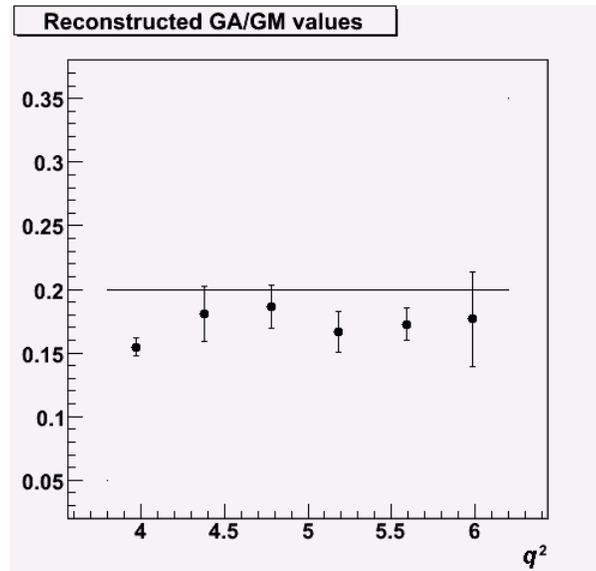}  
\caption{
Extracted $r_a=\vert G_A/G_M\vert$ values from the event distributions of 
Fig.~\protect{\ref{fit11200}} according to the Dip$2\gamma$ parametrization; solid line
is the expected result (see text).}
\label{AM11200}
\end{figure}
%%%%%%%%%%%%%

In Fig.~\ref{AM11200}, the $r_a$ values, reconstructed from Eq.~(\ref{eq:bparam}),
are shown together with the relative errors $\Delta r_a$ from
Eq.~(\ref{eq:err_prop}). The solid line is the expected value $r_a =0.2$. From the
figure it is evident that the error bars (fitting errors) and the discrepancy from
the expected value (statistical errors) are of magnitude $2\Delta r_a \sim 0.05$. 
Therefore, with the considered sample and binning we would be able to distinguish 
a non-Born contribution when it is of the size $\vert r_a\vert > 0.05$.

The additional non-Born corrections to $G_E$ and $G_M$ can be 
effectively reabsorbed in the measurement of the left-right asymmetry in the 
$\cos\theta$ distribution. As shown in Ref.~\cite{2gamma3}, charge conjugation 
invariance imposes general symmetry properties of the Born and $2\gamma$ 
amplitudes with respect to the $\cos\theta\rightarrow-\cos\theta$ transformation.
In particular, the $2\gamma$ corrections to the form factors should respect the 
following constraints
\begin{equation}
\Delta G_{E,M}(q^2,\cos\theta)=-\Delta G_{E,M}(q^2,-\cos\theta) \; ,
\qquad G_A(q^2,\cos\theta)=G_A(q^2,-\cos\theta) \; .
\label{eq:chargeconjg}
\end{equation}
It follows that the contribution of the $\mathrm{Born}\otimes2\gamma$ 
interference term to the unpolarized cross section has the general expression
\begin{equation}
\frac{d\sigma^{\mathrm{(int)}}}{d\cos\theta}=\cos\theta \, [c_0(q^2)\ +\ 
c_1(q^2)\, \cos^2\theta \ +\ c_2(q^2)\, \cos^4\theta\ +\  \dots ]\; ,
\end{equation}
where $c_i$ $(i=0,1,\dots)$  are real coefficients incorporating effects from all 
the three different form factors. In our analysis, where we take 
$\Delta G_{E,M}=0$ and we consider only the first term in the power expansion in 
$\cos\theta$ of $G_A$, the left-right asymmetry proportional to $r_a$ gives us 
only a lower bound for the actual strength of the $2\gamma$ effects. It is 
however evident that several independent observables, including the polarization 
of the recoil proton and/or electron beam, are necessary to disentangle the 
two-photon contribution from each of the three form factors.

%%%UUUUUUUUUUUUUUUUUUUUUUUUUUUUUUUUUUUUUUUUUUUUUUUUUUUUUU

\section{Conclusions}
\label{sec:end}

We have performed numerical simulations of the $e^+ e^- \to p\bar{p}$ 
unpolarized process using a sample of $300\,000$ events in the kinematical region 
with $3.8\leq q^2 \leq 6.2$ GeV$^2$, distributed over 6 equally spaced bins.
For each $q^2$ bin, events were further distributed over 7 equally spaced  
bins in $\cos\theta$, with $\vert \cos\theta\vert<0.9$. For each $q^2$ value, the   
distribution in $\cos\theta$ was fitted by the function 
$A[1\ -\ B \cos\theta\ +\ R \cos^2\theta]$, where the
fit parameters $R$ and $B$ allow for the reconstruction of the 
underlying values of the ratio $r_e \equiv \vert G_E/G_M \vert$ 
and $r_a \equiv \vert G_A/G_M \vert$, respectively, once models for the proton 
form factors  are inserted as input.

Concerning $r_e$, the reconstruction seems to be relatively simple if 
$G_E \sim G_M$, as we expect in the considered kinematical range. 
The size of the sample reproduces the expected $\vert G_E/G_M \vert$ within 
5-10\%. The worst performance is for $q^2 \approx 6$ GeV$^2$, where the bins are
more scarcely populated. 

The additional contributions related to two-photon exchange diagrams show up 
in a left-right asymmetry of the angular distribution, driven by the parameter
$B$. In our analysis, we modeled such a contribution in terms of the axial form 
factor $G_A$, providing a lower bound for the actual strength of two-photon 
effects. The simulation shows that it is possible to identify and estimate this 
term, if $\vert G_A\vert$ (or equivalently $\vert \Delta G_E\vert $ or
$\vert \Delta G_M\vert$) is larger than 5\% of $\vert G_M\vert $, and 
the relative phases of the form factors do not combine in such a way to produce
severe cancellations. However, a more refined analysis, simultaneously including 
also polarization observables, is needed to better constrain and disentangle 
the two-photon contribution from each of the three form factors.

%%%%%%%%%%%%%%%%%%%%%%%%%%%%%%%%%%%%%%%%%%%%%%%%%%%%%%%%%%%%%%%%%%%%%%%%%%%%

%%%%%%%%%%%%%%%%%%%%%%%%%%%%%%%%%%%%%%%%%%%%%%%%%%%%%%%%%%%%%%%%%%%%%%%%%%%%%%


\begin{thebibliography}{}

\bibitem{gao}
C.E.~Hyde-Wright and K.~De Jager, 
Ann.Rev.Nuc.Part.Sci. {\bf 54}, 217 (2004); \\
H.~Gao, 
Int. J. Mod. Phys. {\bf E12}, 1 (2003) 
[Erratum-{\em ibid} {\bf E12}, 567 (2003)]; \\
I.A.~Qattan {\em et al.} [JLab Hall A], 
Phys. Rev. Lett. {\bf 94}, 142301 (2005).

\bibitem{jlab}
V.~Punjabi {\em et al.} [JLab Hall A], 
Phys. Rev. C{\bf 71}, 055202 (2005) 
[Erratum-{\em ibid} {\bf 71}, 069902 (2005)];\\
O.Gayou {\em et al.} [JLab Hall A], 
Phys. Rev. Lett. {\bf 88}, 092301 (2002);\\
M.K.Jones {\em et al.} [JLab Hall A], 
Phys. Rev. Lett. {\bf 84}, 1398 (2000). 

\bibitem{2gamma1}
A.V.~Afanasev {\em et al.},
Phys. Rev. D{\bf 72}, 013008 (2005).

\bibitem{2gamma2}
P.G.~Blunden, W.~Melnitchouk, and J.A.~Tjon, 
Phys. Rev. C{\bf 72}, 034612 (2005).

\bibitem{2gamma3}
M.P.~Rekalo and E.~Tomasi-Gustafsson, 
Eur. Phys. J. {\bf A22}, 331 (2004).

\bibitem{dubnick}
A.Z.~Dubnickova, S.~Dubnicka, and M.P.~Rekalo, 
Nuovo Cimento A{\bf 109}, 241 (1996).

\bibitem{brodsky1}
S.J.~Brodksy {\em et al.}, 
Phys. Rev. D{\bf 69}, 054022 (2004).

\bibitem{egle1} 
E.~Tomasi-Gustafsson {\em et al.}, 
E. Phys. J. {\bf A24}, 419 (2005).

\bibitem{fenice}
A.~Antonelli {\em et al.} [FENICE], 
Nucl. Phys. B{\bf 517}, 3 (1998).

\bibitem{dr}
P.~Mergell, U.-G.~Meissner, and D.~Drechsel, 
Nucl. Phys. A{\bf 596}, 367 (1996); \\
H.-W.~Hammer, U.-G.~Meissner, and D.~Drechsel, 
Phys. Lett. B{\bf 385}, 343 (1996).

\bibitem{e685-2}
M.~Andreotti {\em et al.} [E685], 
Phys. Lett. B{\bf 559}, 20 (2003).

\bibitem{babar}
B.~Aubert {\em et al.} [BaBar], 
Phys. Rev. D{\bf 73}, 012005 (2006).

\bibitem{lear}
B.~Bardin {\em et al.} [LEAR], 
Nucl. Phys. B{\bf 411}, 3 (1994).

\bibitem{baldini}
R.~Baldini {\em et al.}, 
Eur. Phys. J. {\bf C11}, 709 (1999).

\bibitem{LoI}
N.~Apokov {\em et al.}, 
{\em Measurement of the Nucleon Form Factors in the Time-Like region at DAFNE}, \\
Letter of Intent (October 2005), 
see {\tt http://www.lnf.infn.it/conference/nucleon05/loi$\_$06.pdf}

\bibitem{roadmap}
F.~Ambrosino {\em et al.}, 
{\em Prospects for $e^+ e^-$ physics at Frascati between the $\phi$ and the $\psi$}, 
{\tt hep-ex/0603056}.

\bibitem{next}
A.~Bianconi, B.~Pasquini, and M.~Radici, in preparation.

\bibitem{egle2}
G.I.~Gakh and E.~Tomasi-Gustafsson,
Nucl. Phys. {\bf A771}, 169 (2006).

\bibitem{brodsky}
S.J.~Brodsky and G.P.~Lepage, 
Phys. Rev. D{\bf 24}, 2848 (1981); \\
S.J.~Brodsky and G.R.~Farrar, 
Phys. Rev. D{\bf 11}, 1309 (1975).

\bibitem{iachello}
F.~Iachello, A.D.~Jackson, and A.~Lande, 
Phys. Lett. B{\bf 43}, 191 (1973); \\
F.~Iachello and Q.~Wan, 
Phys. Rev. C{\bf 69}, 055204 (2004).

\bibitem{lomon}
E.L.~Lomon, 
Phys. Rev. C{\bf 66}, 045501 (2002).

\bibitem{pbar1} 
A.~Bianconi {\em et al.}, 
Phys.Lett. B{\bf 483}, 353 (2000); \\
G.~Bendiscioli and D.~Kharzeev, 
Rivista del Nuovo Cimento {\bf 17}, 1 (1994). 

\end{thebibliography}
\end{document}